\documentclass[fleqn,twoside]{article}
\usepackage{espcrc2}

\usepackage{graphicx}
\usepackage[figuresright]{rotating}
\usepackage{epsfig}

\newcommand{\AmS}{{\protect\the\textfont2
  A\kern-.1667em\lower.5ex\hbox{M}\kern-.125emS}}

\title{TAUOLA as tau Monte Carlo
  for future applications
  }

\author{ Z. W\c{a}s\address[MCSD]{CERN 1211 Geneva 23, Switzerland \\ 
and Institute of Nuclear Physics, Polish Academy of Sciences,\\
         ul. Radzikowskiego 152, 31-342 Cracow, Poland}%
        \thanks{  Supported in part by  the Polish Government grant KBN 1 P03 091 27   (years 2004-2006).
 Z.W. partly supported also by
Marie Curie Host Fellowship for the Transfer of Knowledge
Contract No. MTKD-CT-2004-510126.}
and 
P. Golonka  \address[MCSD]{CERN 1211 Geneva 23, Switzerland \\  and 
Faculty of Physics and Applied Computer Science,
AGH University of Science and Technology,\\
al. Mickiewicza 30, 30-059 Cracow, Poland.
}$^{*}$
}
       
\begin{document}

\begin{abstract}
The status of the Monte Carlo programs for the simulation of $\tau$-lepton production and decay in 
high-energy accelerator experiments is reviewed. In particular, the status of the
following packages is discussed: (i) {\tt TAUOLA} for $\tau$-lepton decay, (ii) {\tt PHOTOS} for 
radiative corrections in decays,  (iii) {\tt MC-TESTER} packages for various types of semi-automatic tests, and (iv)  universal
interface of {\tt TAUOLA} for the decay of $\tau$ leptons produced by `any' generator. 
Emphasis is put on recent developments for high-precision tests and extensions of {\tt PHOTOS}. 
Some considerations for the software organization necessary in future applications
for Belle and BaBar will be given; examples of {\tt TAUOLA} universal interface use will only be
listed at the end of the review.

\vspace{2mm}
\centerline{ \it Presented at International workshop on Tau Lepton Physics, TAU04 
Nara, Japan September 14-17,2004}
\vspace{1pc}
\centerline{preprint \hskip 1 cm  CERN-PH-TH/2004-230   \hskip 1 cm  HNINP-V-04-05}
\vspace{1pc}
\end{abstract}

\maketitle

\setcounter{footnote}{0}

\section{Introduction}
The {\tt TAUOLA} package
\cite{Jadach:1990mz,Jezabek:1991qp,Jadach:1993hs,Golonka:2000iu}  for the simulation 
of $\tau$-lepton decays and  
{\tt PHOTOS} \cite{Barberio:1990ms,Barberio:1994qi} for the simulation of radiative corrections
in decays, are computing
projects with a rather long history. Written and maintained by 
well-defined authors, they nonetheless migrated into a wide range
of applications where they became ingredients of 
complicated simulation chains. As a consequence, a large number of
different versions are presently in use.
From the algorithmic point of view, they often
differ only in a few small details, but incorporate many specific results from distinct
$\tau$-lepton measurements. Such versions were mainly maintained (and will remain so) 
by the experiments taking precision data on $\tau$ leptons. On the other hand,
 many new applications were developed  recently,  often requiring
a program interface to other packages  (e.g. generating events for LHC, LC, 
Belle or BaBar physics processes).

In the last two years some progress in  the simulation 
of $\tau$-pair production alone was achieved. A control of the
systematic errors was improved and a special tool developed for that
purpose, {\tt MC-TESTER}, turned out to be very useful. 
With its help, the new version of {\tt TAUOLA}, which may become the starting point for 
the work directly oriented towards phenomenology of b-factories, was prepared.
{\tt MC-TESTER} became instrumental in developing the new version of {\tt PHOTOS}
for radiative corrections in decays. With its help tests enabling 
verification of the program in multiple photon radiation mode became possible,
thus establishing {\tt PHOTOS} as a precision tool for simulation of 
radiative bremsstrahlung in $W$, $Z$ and 
Higgs boson decay. New results for $\tau$ physics were obtained 
as well.  

Because of the limited space and sizeable amount of physically interesting 
numerical results, only some of them will be included in the conference 
proceedings. For the remaining ones we refer the reader to references 
and transparencies from the talk.

\section{Versions of {\tt TAUOLA} Monte Carlo}

In refs. \cite{Golonka:2000iu,Golonka:2003xt} the setup for constructing specific versions of {\tt TAUOLA}
and  {\tt PHOTOS} from the single set of files was prepared and documented. The system  
was  prepared for the software librarians and advanced users interested in updating
the packages for the multipurpose environment. 
The idea was to create a repository which allows one to keep all main 
options of {\tt TAUOLA} developed for different purposes in a relatively compact form,
without duplications of semi-identical parts. The
repository was set to produce standard {\tt FORTRAN} files 
which can later be handled
exactly the same way as the early 1990 versions of the packages,
but with physics initialization based on Cleo and Aleph data of the years
1997 -- 98.
During the conference, another generating system was developed in parallel, to start 
new work with the Belle and BaBar collaborations. Physics initialization remained
as in Cleo, but 60 new channels were prepared 
(without physical initialization) for the future implementations to be done within
those collaborations. 
To facilitate integration, the generic subdirectory {\tt TAUOLA/tauola-BBB} was prepared.
The new versions can be now initialized with the help of the command
{\tt make} with parameters {\tt authors}, {\tt Belle} or {\tt BaBar}.
Whether this solution will become accepted needs to be investigated. The geographical spread
of interested people over three continents is certainly a disadvantage.

\section{{\tt PHOTOS}}
\def\CCol{{\tt SANC}}
Recently some extensive tests and extensions were introduced into {\tt PHOTOS}, an algorithm for radiative corrections 
in decays of particles or resonances. Thanks to comparisons with the single-photon exact matrix-element for the $W$ 
decay
correcting weight was introduced (in this particular case, see fig \ref{NuFig}).
The method applied can also serve as an example for similar correcting 
weights to be introduced in other decay channels, of more importance for the b-factories community. In theses cases
correcting weights may be of a more phenomenological nature, and to some degree may represent a fit to the data.
%
\begin{figure}[!ht]
\centering
\setlength{\unitlength}{0.1mm}
\begin{picture}(800,1200)
\put( 375,750){\makebox(0,0)[b]{\large }}
\put(1225,750){\makebox(0,0)[b]{\large }}
\put(-20, -100){\makebox(0,0)[lb]{\epsfig{file=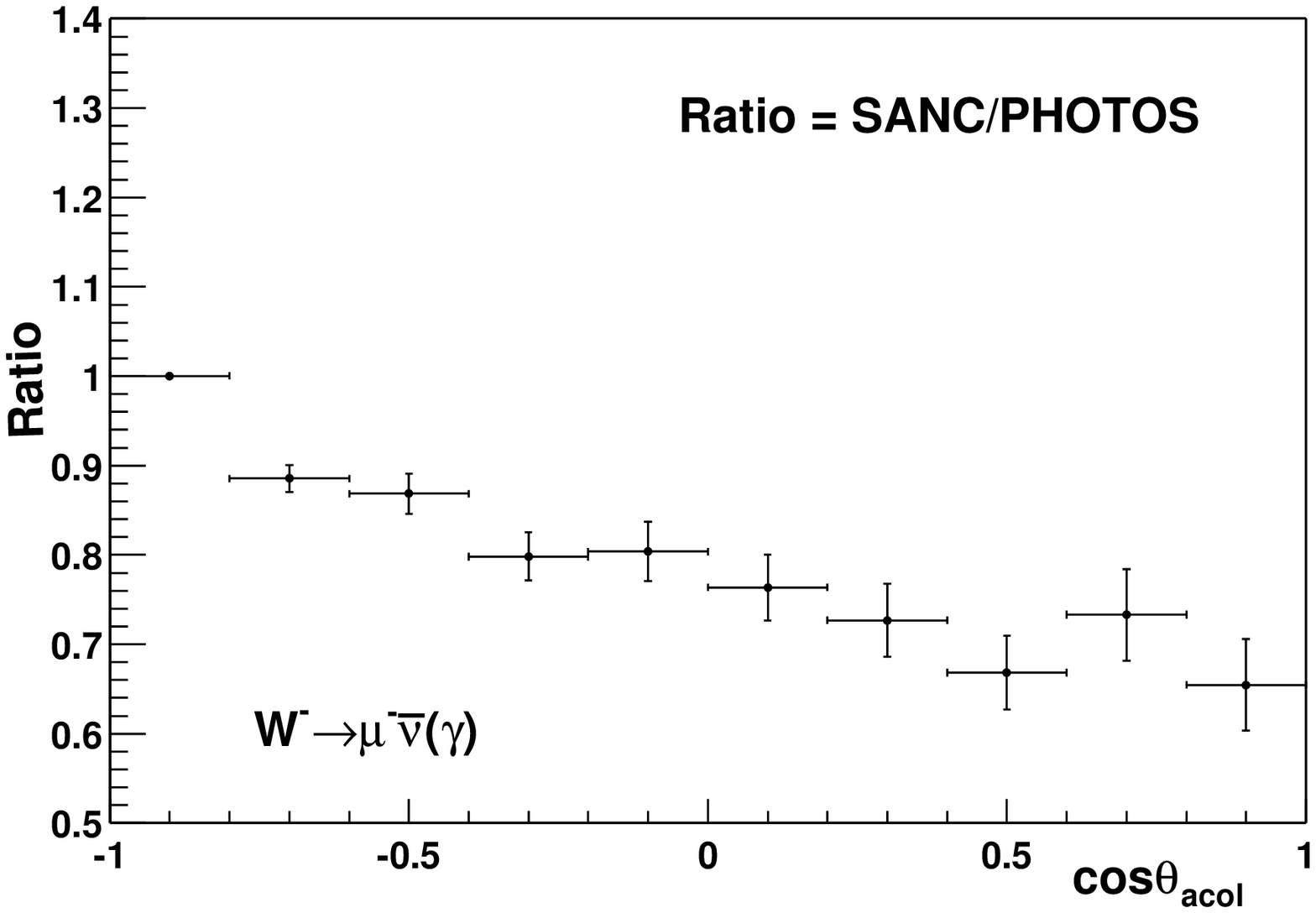,width=80mm,height=40mm}}}
\put(-20, 340){\makebox(0,0)[lb]{\epsfig{file=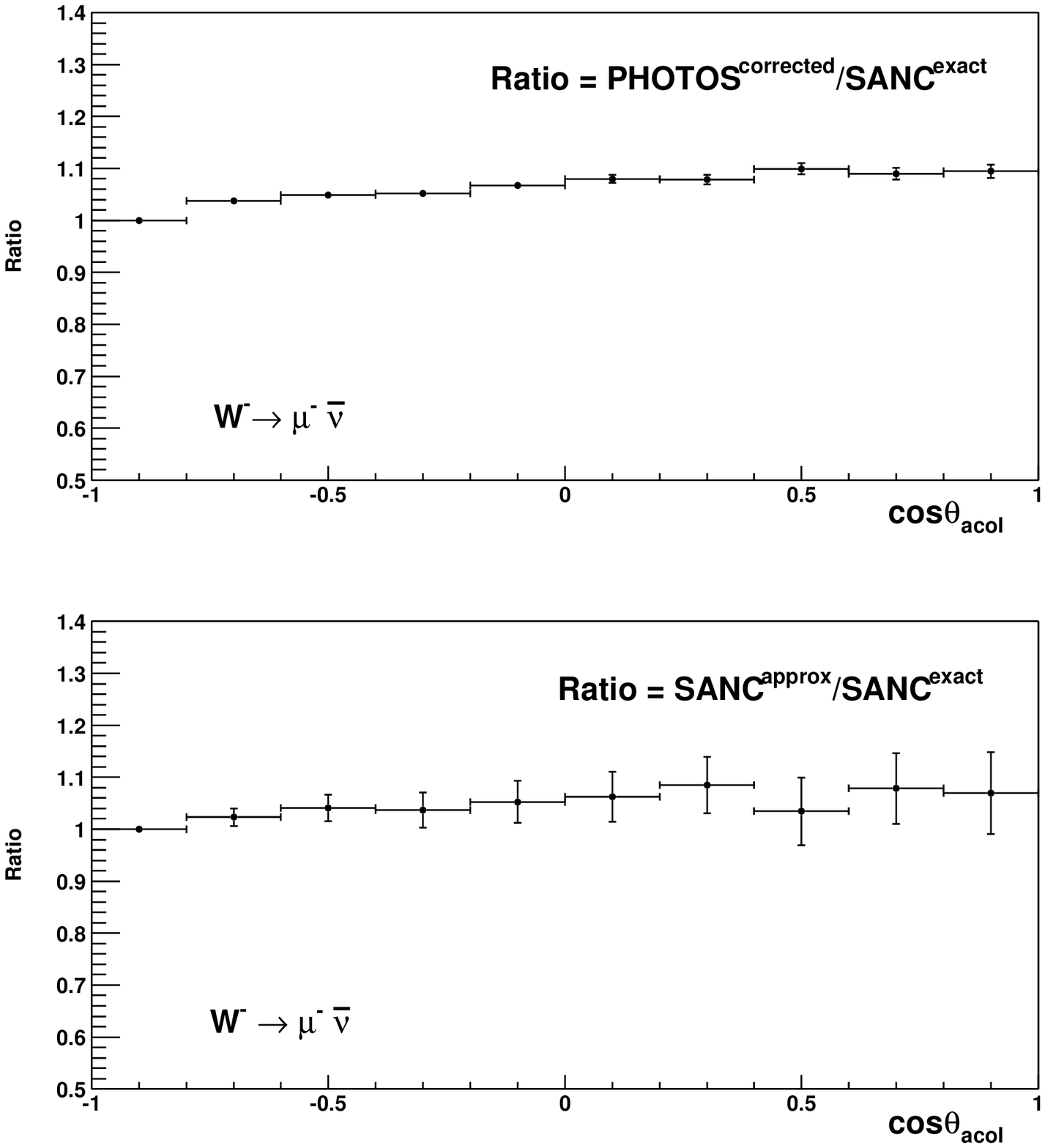,width=80mm,height=80mm}}}
\end{picture}
\caption[]{\small\sf Comparisons (ratios) of the \CCol\, and {\tt PHOTOS} predictions 
for the $W$ decay, distribution of $\mu^{-} \nu$ acolinearity angle, $\cos(\Theta_{acol})$ is given.
Ratio of the distributions from the two programs is plotted. 
The dominant contribution to the distribution is from the non-infrared, non-leading-log part of the 
phase-space, i.e. the region which is most sensitive to non-leading contributions \\
{\bf Upper two plots:} Distribution with correction weight, plot taken from  ref. \cite{Nanava:2003cg}. Results with approximation
installed in  \CCol\ matrix element provide additional technical test.\\
{\bf Lower plot:} Distribution without correction weight plot taken from ref. \cite{Andonov:2002mx} 
}
\label{NuFig}
\end{figure}

In the second part of the extension of {\tt PHOTOS} we have followed the solution presented 
in ref. \cite{Barberio:1994qi} for the iterative algorithm of multiple photon emission. 
Thanks to a better understanding of the numerical stability of the algorithm, we were
able to go beyond double photon-emission, and extend the iteration up to quadruple emission. Once 
this was achieved we were able to complete the exponentiation programme, still preserving full control of terms
providing leading logs as well as the soft photon region of the photon emission. It is important to note that the algorithm
 covers  full phase space for photon emission, except soft regions where photon emission is integrated analytically and 
added together to the virtual corrections. 
Thanks to the exponentiation, we could lower the value of the minimal energy of generated photons
to the level of $10^{-5}$ or $10^{-6}$ without any problems. This was plaguing the project until
now.

This goal required an important development for numerical stability. {\tt PHOTOS} operates on four-vectors, that is why explicit
generation of photons at small energies immediately leads to loss of the numerical stability of additional 6 orders.
 Therefore a special algorithm to prevent accumulation of rounding errors was developed.
On the technical side let us remark that this special correcting routine, {\tt PHCORK},
had to be called after every emission of individual photons (routine {\tt PHOMAK}. 
{\tt PHOMAK} has to assure that energy-momentum is conserved in 
the decay branch and that four-momenta of all particles are precisely on mass-shell. 

Once the list of necessary modifications was completed we performed extensive tests using the method presented in the next section.
This helped to establish the precision of {\tt PHOTOS} in the case of $Z$ decay at the permille level. Comparison tests
with {\tt KORALZ} and {\tt KKMC} support such conjecture. In the case of $W$ decay it was confirmed by comparison with {\tt WINHAC}
\cite{Placzek:2003zg}.

\section{ {\tt MC-TESTER} and its extensions}
{\tt MC-TESTER} is described in detail in ref.~\cite{Golonka:2002rz}; the principle of this testing package is to analyse series
of events originating from a Monte Carlo generator (or a system of several Monte Carlos combined together). 
In fact only a branch originating from the decay of some intermediate resonance or particles is examined.
In a fully automated way (thereby limiting manpower and also the risk of accidental errors) the output from 
the run is created. If a similar run for a different generator (possibly even written in different programing language) is 
performed, then the two files with information collected by {\tt MC-TESTER} can be compared and the result of the test can be
presented in a visualised form as a booklet of plots and tables with general information on the run, a list of the decay channels,
their branching ratios, and {\tt SDP}\footnote{Shape Difference Parameter: quantifies how much the shape
of the histograms (of all possible invariant masses which can be constructed from the momenta of the decay products 
of the particle under study) differs between the two runs of distinct programs.} parameter.

Such a test was very useful in the case of validation of {\tt TAUOLA}. However, in cases where an arbitrary number of final-state particles 
could be present (as in the case of final-state QED bremsstrahlung) the test required an extension.

 There may be  quite big differences in the way bremsstrahlung is generated by the programs for decays. 
In our test, we aim to develop a technique where comparisons make physical sense and are automatic. 
To this end, we have to free our test from dependence on the technical parameters of the generators, 
such as a minimal photon energy threshold, hence ambiguities due to the number of soft photons generated, explicit choices for
Lorentz frames of generation, etc.

Independently from how events are generated, we define zero-, one- or two-photon topologies.
We will call the event to be of `zero photons', if there is no photon at all of energies larger than the $E_{test}$ parameter 
of the test. The `one-photon' event will have to have one (and only one) photon of energy larger than $E_{test}$. 
If there is more than one such photon, we shall call it `two-photon' event. If there are more than 
two photons of energy larger than  $E_{test}$, then we consider only the two most energetic photons, and treat the remaining 
ones as if they did not pass the  $E_{test}$ threshold.

There may be cases where more than two photons actually considered (e.g. with energies not passing the  $E_{test}$ threshold) 
are present.
Following the leading-log-inspired logic, we will add them to the momenta of outgoing fermions (always to the fermion 
of a smaller angular separation).

We define two tests: {\tt test1} and {\tt test2}.  {\tt Test2} is exactly as explained above.
In case of {\tt test1}, only one photon (the most energetic one) can be accepted.

The test has therefore two parameters:

-- maximal multiplicity (of photons) defines whether {\tt test1} or {\tt test2} is used

--  minimum photon energy:  $E_{test}$

\vskip 3 mm
{\bf Tests of {\tt PHOTOS} in $W$ and $Z$ decays}
\vskip 3 mm

\begin{figure}[!ht]
\centering
\setlength{\unitlength}{0.1mm}
\begin{picture}(800,1500)
\put(-20, 710){\makebox(0,0)[lb]{\epsfig{file=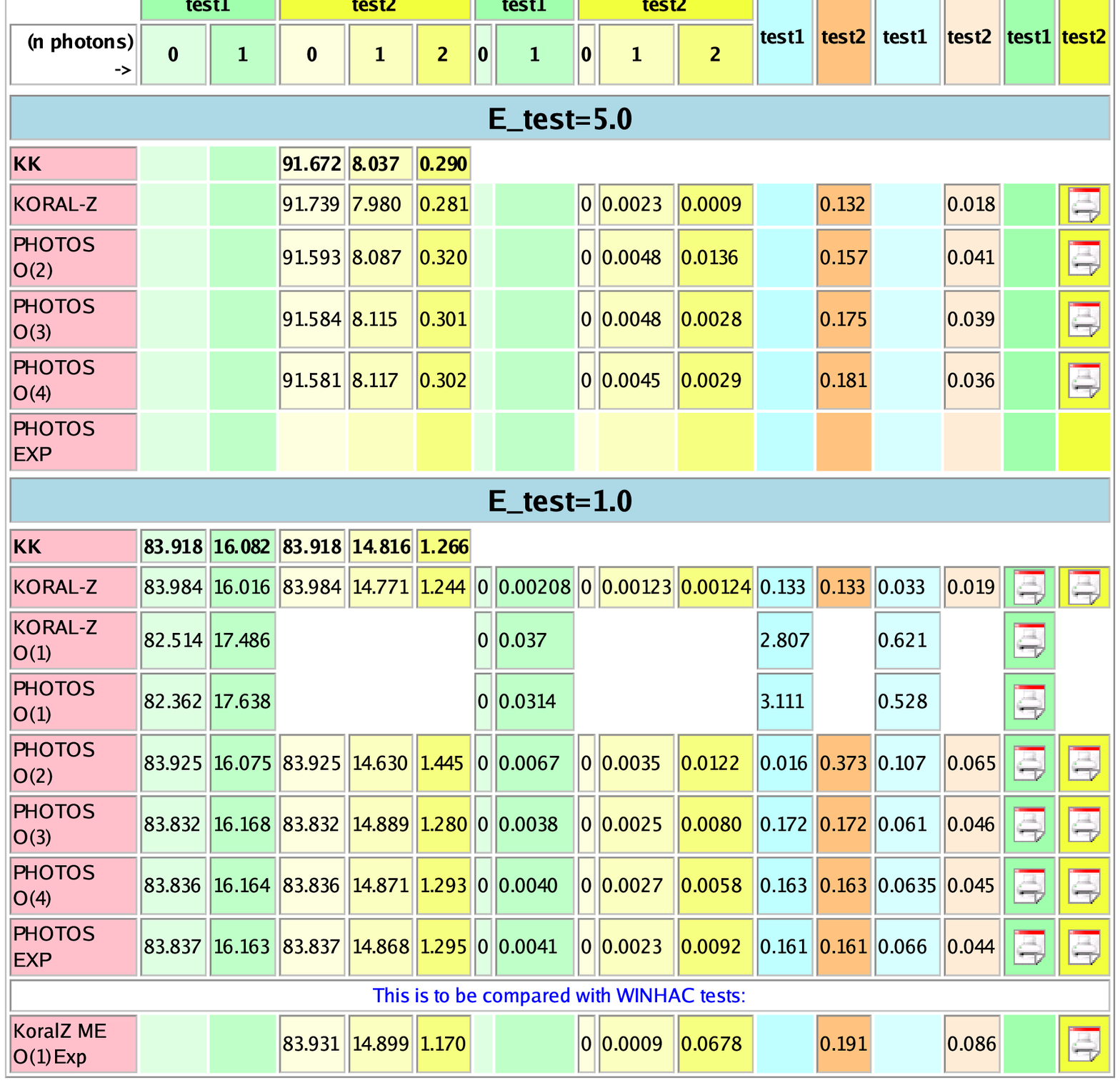,width=80mm,height=90mm}}}
\put(-20,  400){\makebox(0,0)[lb]{\epsfig{file=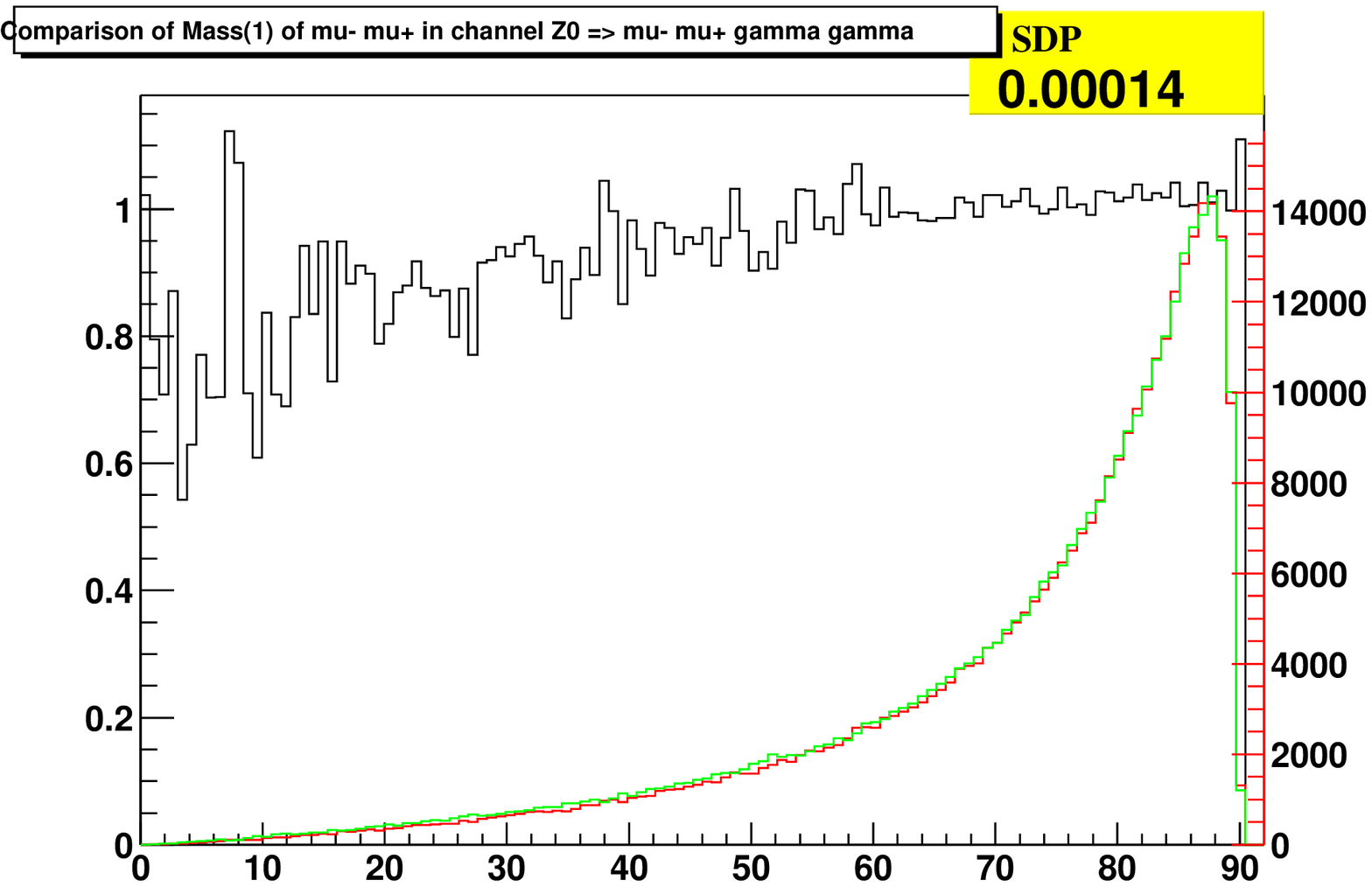,width=80mm,height=50mm}}}
\put(-20, -120){\makebox(0,0)[lb]{\epsfig{file=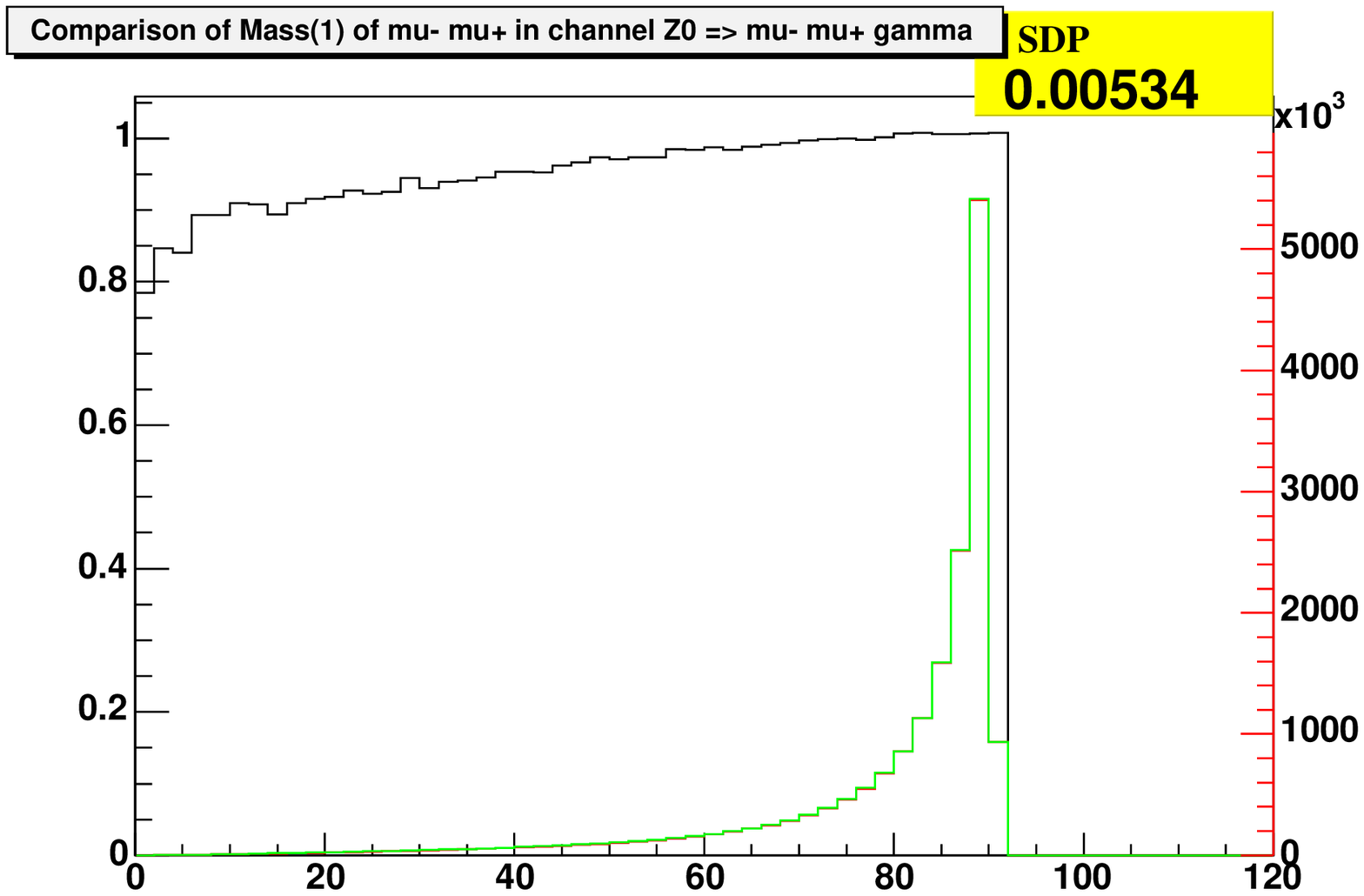,width=80mm,height=50mm}}}
\end{picture}
\caption{ \small \sf {\bf Upper part}:  Comparison 
table for  {\tt PHOTOS},  {\tt KKMC} and {\tt KORALZ} results. For more details see the text.
{\bf Middle plot}:
Single plot from the comparison booklet. 
 The chosen plot presents the distribution of the $\mu^+ \mu^-$ invariant mass. There
the largest discrepancy   between {\tt PHOTOS EXP}
and {\tt KKMC} was found. 
{\bf  Bottom plot: }
The same difference in case of single photon  runs  of {\tt PHOTOS} and {\tt KORALZ} and $\mu^+ \mu^-$ invariant mass
can be seen. One can see that the source 
of residual difference is in approximation of the matrix-element used in {\tt PHOTOS} and shows the possible 
further improvement in the code. }
\label{fig:ifi-second}
\end{figure}

Tests of comparisons for {\tt PHOTOS} with {\tt KKMC} and {\tt KORALZ} \cite{kkcpc:1999,koralz4:1999} for $Z$ decay and with 
{\tt WINHAC} \cite{Placzek:2003zg} for $W$
are collected on the webpage ~\cite{tauolaphotos}. These tests are quite extensive and we can not 
present here even the most important results. That is why only an example illustrating the general form of the 
results is given in fig.~\ref{fig:ifi-second}. The upper part of the figure
presents the comparison `hyper-table'. Comparisons of {\tt PHOTOS} running with different options 
(single- double- triple- quadruple- and multiple-radiation with the results from
 {\tt KKMC} and {\tt KORALZ} Monte Carlo programs  are collected.
The first column of the table defines the generator used (with options) for the numerical
results collected; the second part of the table collect
branching ratios obtained from {\tt test1} and {\tt test2} as described above,
respectively for configuration with 0,1 ({\tt test1}) and 0,1,2 photons 
({\tt test2}). Later the maximum of the {\tt SDP} (as defined in ref. \cite{Golonka:2002rz})
 is given. As reference point, the first
entry after the definition of $E_{test}$ is used; in this table
results from  {\tt KKMC} are used. 
The last two columns represent hyper-links to the comparison booklets 
(not included in the articles but available from \cite{tauolaphotos}). 
The booklets of standard {\tt MC-TESTER} format \cite{Golonka:2002rz}
include the plots like the one presented in the middle of fig.~\ref{fig:ifi-second}
for invariant mass of $\mu^+\mu^-$ where the largest difference between results
from the {\tt KKMC} and {\tt PHOTOS} version with exponentiation is visible. 

One can observe, rather unexpectedly, a good agreement between {\tt PHOTOS} and {\tt KKMC}. In fact only in the case when 
the full double-photon matrix-element in {\tt KKMC} is used is the agreement with {\tt PHOTOS} good. 
If a less powerful matrix-element that does not rely on the spin amplitude exponentiation
scheme of {\tt KORALZ} is used, or a second-order matrix-element is switched off 
in {\tt KKMC}, differences from the full exponentiated
version of {\tt PHOTOS},  quantified with the help of SDP, 
increase. 
What could be even more unexpected is that the effects of coherence between photons is to a large 
degree reproduced by the {\tt PHOTOS} algorithm. This delicate point is still under investigation. 
Some numerical results are collected 
on the web page \cite{tauolaphotos} in subsection `PHOTOS improved antenna algorithm'. 
However, this point is of no 
numerical interest for bremsstrahlung in decays, at least not for present-day applications, because the observed effects are too small.
On the other hand, this point  is interesting as a potentially important ingredient for constructing 
high-precision Monte Carlos
for future applications; either for QED at linear colliders, or for QCD at the LHC.

In the case of $W$ decays our comparisons for multiple photon emission were limited to comparisons with {\tt WINHAC} \cite{Placzek:2003zg}.
Less good agreement, but still better than required by present-day applications, was achieved. We did not investigate
 the sources of the residual discrepancies. The good candidates were limitations of the correcting weight used in {\tt PHOTOS},
 see discussion above, and the fact that the {\tt WINHAC} matrix-element is limited to the first order only. More detailed numerical discussion
 of this point could not be included in the conference contribution presented because of space limitation.

\vskip 3 mm
{\bf Tests of {\tt TAUOLA} and {\tt PHOTOS} in leptonic $\tau$ decays}
\vskip 3 mm
In the case of the $\tau$ lepton physics, the discussion presented above may be of no practical consequences. However, the methodology 
developed for QED bremsstrahlung (and tests) in case of $W$ and $Z$ decays could also be applied for leptonic 
$\tau$ decays. With the methods analogous as presented above, we have found that the difference
between single-photon full matrix-element generation of {\tt TAUOLA} and the single-photon option of {\tt PHOTOS} is larger 
than between the first-order and exponentiated mode of operation of {\tt PHOTOS}, see ref. \cite{tauolaphotos}. We could conclude
that the matrix element of {\tt TAUOLA} for $\tau \to \nu \bar \nu \mu (\gamma)$ is better than the exponentiated version of {\tt PHOTOS}.
This conjecture is true at least for semi-inclusive quantities as discussed here. Of course for generation of events
with explicit multiple-photon final-states, the results generated with {\tt PHOTOS} will be more precise. 

\section{Universal interface based on {\tt HEPEVT} common block}

The universal interface of {\tt TAUOLA} for `any' $\tau$ production generator
is distributed together with {\tt TAUOLA} \cite{web-page,Golonka:2003xt}.
 It uses as an input the {\tt HEPEVT} common block and operates on its content
only. As a demonstration example the
interface is combined  with the {\tt JETSET}  generator \cite{jetset6.3:1987}.

The interface works in the following way:
\begin{itemize}
\item
The $\tau$-lepton should be forced to be stable in the $\tau$ production  generator.
\item
The content of the {\tt HEPEVT} common block is searched for all $\tau$
leptons and neutrinos first. 
\item
We check if there are $\tau$-flavour pairs (two $\tau$-leptons or
$\tau$-lepton and $\tau$-neutrino) originating from the same mother. 
\item
The decays of the $\tau$-flavour pairs are performed with {\tt TAUOLA}.
Longitudinal spin effects are generated  in the case of the $\tau$ produced from the decay of:
$W \to \tau \nu$, $Z/\gamma \to \tau \tau$, 
the neutral Higgs boson $H \to \tau \tau$, and the charged Higgs boson
 $H^{\pm}\to  \tau \nu$.
Parallel or anti-parallel spin configurations are
generated, before calling on the $\tau$ decay, and then the decays of 100\% polarized $\tau$'s
are executed. 
\item
In the case of the Higgs boson (for the spin-correlations to be generated)
the identifier of the $\tau$'s mother must be that of the Higgs. In this case
it was rather easy to implement full spin-correlations.
\item
In case the $\tau$ lepton is not produced from any of the above-mentioned
intermediate states, then if there is produced alsoa $\nu_\tau$ from the same 
mother as that of the $\tau$ , the $W$ is reconstructed by the interface as the sum of the two. 
Similarly, $Z/\gamma$ is reconstructed if  another $\tau$ is produced from the
same mother.

\end{itemize}

This organization of software became quite productive. It was helpful to 
perform studies of observables for Higgs boson parity measurement at future 
LCs. In this case it was important that full spin correlations for the 
multi-layer cascade
process $h \to \tau^+ \tau^-$, $\tau^\pm \to \nu \rho^\pm$, 
$\rho^\pm \to \pi^\pm \pi^0$ be included \cite{Was:2002gv,Bower:2002zx,Desch:2003mw,Desch:2003rw}.
Also, in the case of studies for the Higgs discovery potential at LC, the implementation of 
the universal interface was useful and productive \cite{Richter-Was:2004jf} 
to better visualize possible difficulties. Recently \cite{Chankowski},
the interface had proven to be useful for studies of CP parity effects in 
$B_0 \bar B_0$ system produced at Belle/BaBar energies, when one of the $B$-mesons
decays into a pair of $\tau$ leptons. This process, even though not discovered
so far, provides another example of the usefulness of the universal interface.
%

\section{Summary and future possibilities}

The status of the computer programs for the decay of $\tau$ leptons and
associated projects  was reviewed.
The high-precision version of  {\tt PHOTOS} for radiative corrections was presented.
 In particular, the option to run the program with multiple-photon radiation
was presented. The results for  decays of $\tau$-leptons, $Z$ and $W$ bosons were mentioned.
It was found that for semi-inclusive single- and double-photon final-states the agreement
with the matrix element simulations of the second-order is at the 0.1\% level. Only
at the tails  of some distributions, the approximation 
used in {\tt PHOTOS} was visible
and gave a discrepancy of the 15\% level in  sparsely populated corners of the phase-space.

 The presentation of the {\tt TAUOLA} general-purpose interface
  was ommited because of lack of time. Examples for its use in the case of the Higgs boson parity
measurement at a future Linear Collider \cite{Bower:2002zx,Desch:2003mw,Desch:2003rw} 
and for Higgs searches at the LHC can be found 
in the literature.  Recently, a similar application was  developed for the
case of studies in hypothetical effects of  CP-parity breaking in the $B_0$-$\bar B_0$ system
at Belle and BaBar \cite{Chankowski}.
At present, in its full content, it is available from the authors upon individual request only.

Distinct versions of the {\tt TAUOLA} library for $\tau$ lepton decay, and of
{\tt PHOTOS} for radiative corrections in decays, are now in use. The principles of the distribution
package, as presented in ref. \cite{Golonka:2003xt}, are carefully preserved
to save continuity of the
project, and for the comfort of the users. In this scheme the new version of {\tt PHOTOS} is now available
from the  home page of Z. Was \cite{web-page}.  However, on top of the previously available versions
of {\tt TAUOLA}, an extension was prepared to introduce
new optional and/or missing decay modes from the independent files. 
In these generating files  60 new channels of $\tau$ decays were drafted
on top of the previously available ones. As previously,
the system for creating the required version of the {\tt PHOTOS}
and {\tt TAUOLA} packages from the single  master copies is preserved. The master copies 
are kept
in relatively compact and clear  form, without unnecessary code duplications.

Development of {\tt MC-TESTER} was a necessary step to prepare the translation of {\tt TAUOLA} to new programming language
such as {\tt C++}, however, it is also very important to perform such transformation 
together with the program's users. At present, there is no pressure from the experimental 
community for such transformation to be imminent. As far as the package authors are concerned,
such a possibility has been envisaged for some time now.

\vskip 3 mm
\centerline{ \bf Acknowledgements}
\vskip 3 mm
 Useful collaboration  and suggestions from the {\tt TAUOLA} and related programs co-authors:
 S. Jadach, J. H. K\"uhn, S. Eidelman, B. Kersevan, T. Pierzchala, E. Richter-Was
and M. Worek are acknowledged. Discussions with  
A. Weinstein and members of the Belle and BaBar collaborations 
are also acknowledged.

\providecommand{\href}[2]{#2}


\end{document}